\documentclass[
reprint,
superscriptaddress,
amsmath,amssymb,
prx,
floatfix,
longbibliography
]{revtex4-2}

\usepackage{graphicx}
\usepackage{dcolumn}
\usepackage{bm}
\usepackage[T1]{fontenc} 
\usepackage{color,soul}
\usepackage{float}
\usepackage{natbib}
\usepackage{setspace}
\usepackage{soul}
\usepackage{color}

\begin{document}

\preprint{APS/123-QED}

\title{Threshold and laser-conversion in nanostructured-resonator parametric oscillators}

\author{Haixin Liu}
    \email[Correspondence email address: ]{haixin.liu@colorado.edu}
     \affiliation{Time and Frequency Division, National Institute of Standards and Technology, Boulder, CO, USA}
     \affiliation{Department of Physics, University of Colorado, Boulder, CO, USA}
\author{Grant M. Brodnik}
     \affiliation{Time and Frequency Division, National Institute of Standards and Technology, Boulder, CO, USA}
     \affiliation{Department of Physics, University of Colorado, Boulder, CO, USA}
\author{Jizhao Zang}
     \affiliation{Time and Frequency Division, National Institute of Standards and Technology, Boulder, CO, USA}
     \affiliation{Department of Physics, University of Colorado, Boulder, CO, USA}
\author{David R. Carlson}
     \affiliation{Time and Frequency Division, National Institute of Standards and Technology, Boulder, CO, USA}
     \affiliation{Octave Photonics, Louisville, CO, USA}
\author{Jennifer A. Black}
     \affiliation{Time and Frequency Division, National Institute of Standards and Technology, Boulder, CO, USA}
\author{Scott B. Papp}
     \affiliation{Time and Frequency Division, National Institute of Standards and Technology, Boulder, CO, USA}
     \affiliation{Department of Physics, University of Colorado, Boulder, CO, USA}

\date{\today}

\begin{abstract} 
We explore optical parametric oscillation (OPO) in nanophotonic resonators, enabling arbitrary, nonlinear phase-matching and nearly lossless control of energy conversion. Such pristine OPO laser converters are determined by nonlinear light-matter interactions, making them both technologically flexible and broadly reconfigurable. 
We utilize a nanostructured inner-wall modulation in the resonator to achieve 
universal phase-matching for OPO-laser conversion, but coherent backscattering also induces a counterpropagating pump laser. This depletes the intra-resonator optical power in either direction, increasing the OPO threshold power and limiting 
laser-conversion efficiency, the ratio of optical power in target signal and idler frequencies to the pump. 
We develop an analytical model of this system that emphasizes an understanding of optimal laser conversion and threshold behaviors, and we use the model to guide experiments with nanostructured-resonator OPO laser-conversion circuits, fully integrated on chip and unlimited by group-velocity dispersion.
Our work demonstrates the fundamental connection between OPO laser-conversion efficiency and the resonator coupling rate, subject to the relative phase and power of counterpropagating pump fields. We achieve $(40\pm4)$ mW of on-chip power, corresponding to $(41\pm4)$\% conversion efficiency, and discover a path toward near-unity OPO laser conversion efficiency.

\end{abstract}

\maketitle

\paragraph*{Introduction.}



Optical parametric oscillation (OPO) features behaviors that are observed in many physical systems. The intensity distribution of the optical field shows a kind of Turing pattern, which is similar to those in biological systems \cite{bullara2015pigment} and sand dunes \cite{kroy2002minimal}. The patterns in these systems arise from nonlinearity of the reaction–diffusion equation. With sand dunes there is a nonlinear surface velocity profile \cite{ amarouchene2001dynamic} while for nonlinear optics it is the nonlinear refractive index. Since OPO is a coherent process, it is subject to a narrow range of phase-matching solutions. Still, we can trace the coherent oscillating output of the OPO to its constituent nonlinear dynamics \cite{kippenberg2018dissipative, yu2021spontaneous}. For example, accidental \cite{xue2015mode} and controllable 
\cite{lu2014selective, xue2015normal, miller2015tunable} mode frequency shifts have been used to balance group-velocity dispersion (GVD) and Kerr shifts in microresonator OPOs. More recently, photonic-crystal resonators (PhCR) have provided a route to phase-match OPO in nearly any dispersion regime \cite{black2022optical}. The accessibility of these controls for nonlinear dynamics in OPO makes the system both interesting to search for novel phenomena and to understand similar dynamics in related physical systems.




OPO laser conversion also has numerous applications in engineering. Degenerate OPO works as a converter of the pump-laser frequency to a tunable signal and idler frequency, which provides a coherent source with designable wavelength \cite{lu2019milliwatt}. With the help of microresonators, OPO laser-converters can be chip integrated. An intrinsic condition of OPO is phase-matching, which is traditionally achieved by designing anomalous GVD to balance Kerr frequency shifts. With GVD engineering, OPO in microresonators has been realized in silica \cite{kippenberg2004kerr,sayson2017widely}, aluminum nitride \cite{tang2020widely}, silicon nitride \cite{lu2020chip,domeneguetti2021parametric}, and tantalum pentoxide \cite{jung2021tantala} platforms. 
Microresonator OPO has also been explored with novel bound states in the continuum to tailor wavelength-dependent coupling conditions \cite{lei2023hyperparametric}.
With PhCR OPOs \cite{black2022optical}, nanostructuring the microresonator waveguide induces coherent backscattering and a controllable frequency splitting of one (or more) azimuthal modes. This provides for direct phase-matching between three modes of the device. However, backscattering the pump also depletes the available power in either counterpropagating direction.

While access to tunable phase matching is important for a laser converter, the conversion efficiency (CE) defined as the ratio between the signal and idler output power and the pump laser power is essential as well. 
According to previous research, the highest CE of microresonator OPO 
with standard couplers is <40\%, and the highest reported on-chip output is $\approx$ 20 mW \cite{kippenberg2004kerr, sayson2019octave, zhou2022hybrid, stone2022efficient, perez2023high}. 
In the case of a 
PhCR, backscattering reduces the power available for OPO threshold and conversion efficiency in a single propagation direction. Despite this, CE in PhCRs exceeding 10\% has been demonstrated by operating in the over-coupled regime \cite{black2022optical}. On the other hand, for frequency-comb generation, which has a similar operating principle as OPO \cite{coillet2013azimuthal, pasquazi2018micro,yu2022continuum}, a pump-to-comb conversion efficiency as high as 83\% has recently been demonstrated by placing a pump reflector to maximize the pump intensity in a PhCR \cite{yu2019photonic,zang2022near}. However, the upper limit of CE and optimal system integration of the PhCR with pump reflector remains an open research question. 
Moreover, a physical understanding of this system will enable future devices with access to large parametric gain for broadly reconfigurable wavelength access and the highest output power. Such chip-scale technologies would be useful for applications as diverse as optical telecommunications \cite{fulop2018high,brodnik2021optically}, spectroscopy \cite{suh2016microresonator}, and optical sensors \cite{lai2020earth, bao2021architecture}. 

Here, we develop an analytical framework to describe OPO laser converters with counterpropagating pump fields, and we derive formulas for CE and threshold power. 
Moreover, we develop and implement the experimental infrastructure for an integrated PhCR with a pump reflector in the bus waveguide.
The PhCR induces coherent backscattering within the resonator, and the pump reflector transforms the pump into counterpropagating fields. 
Thereby, we control the phase between the counterpropagating pump fields in the bus waveguide and the backscattered pump mode inside the PhCR, enabling suppression of the unused pump power in one direction and realization of the optimal utility of pump power.
Our experiments systematically explore the interaction of counterpropagating pump laser fields in the PhCR, which we find to be in good agreement with our analytical model of the nonlinear system. 
 We demonstrate $(41\pm4)$\% CE of OPO by measuring the output power, which is the sum of the power of idler and signal in both directions, and we calculate CE
 as the ratio between the on-chip output power and the on-chip pump power. This measurement corresponds to $(40\pm4)$ mW on-chip output power with idler and signal wavelengths at $(1620.0\pm0.4)$ nm and $(1498.7\pm0.4)$nm. The corresponding spectra are in Supplemental Material. This work illuminates a regime of OPO in which we unleash universal phase matching and high CE through nanophotonic design.


\paragraph*{Theory.}

To derive the CE formula in the pump-reflected PhCR case, we need to understand the dynamics of the system, which are two-fold: a linear process describing the coupling between the bus waveguide and the resonator and a nonlinear process within the resonator due to resonant field enhancement. The nonlinear process is described by the normalized, modified Lugiato-Lefever Equation (LLE) \cite{skryabin2020hierarchy}:
   
\begin{equation}
\begin{split}
\frac{\partial E_{t\mu}}{\partial t} = &-(1+i(\alpha+D_\mu))E_{t\mu} \\&+ i(\sum_{\mu_1,\mu_2}E_{t\mu_1}E_{t\mu_2}E_{t(\mu_1+\mu_2-\mu)}^*+2E_{t\mu}\sum_{\mu_3}I_{r\mu_3}) \\&+ (F_t-i\frac{\xi}{2}E_{r\mu})\delta_{\mu,0}\\
\frac{\partial E_{r\mu}}{\partial t} = &-(1+i(\alpha+D_\mu))E_{r\mu} \\&+ i(\sum_{\mu_1,\mu_2}E_{r\mu_1}E_{r\mu_2}E_{r(\mu_1+\mu_2-\mu)}^*+2E_{r\mu}\sum_{\mu_3}I_{t\mu_3}) \\&+ (F_r-i\frac{\xi}{2}E_{t\mu})\delta_{\mu,0}.
\end{split}
\label{LLE}
\end{equation}

The diagram in Fig. 1(a) shows the components of the field in our devices. The fields in the bus waveguide propagate in both transmitted and reflected directions, denoted by subscript $t$ and $r$, respectively. The fields in the PhCR also have two counterpropagating directions: clockwise (CW) and counterclockwise (CCW). The CW wave in the ring couples with the transmitted direction light in the bus waveguide, while the CCW wave in the ring will couple with the light in the reflected direction within the bus waveguide. Due to the periodic boundary of the microresonator, the field inside a PhCR can be decomposed into different modes in each direction, denoted by $E_{t\mu}$ (CW) and $E_{r\mu}$ (CCW) where $\mu$ represents the mode number relative to the pump mode, $\mu = 0$. The field here is normalized such that the nonlinearity thresholds when $E_{t\mu} (E_{r\mu}) \sim 1$. The intensity of each mode is denoted by $I_{t\mu}$ and $I_{r\mu}$, equal to the square norm of the field. The parameters $F_t$ and $F_r$ represent the effective driving force inside the resonator of both transmitted and reflected directions, $\delta_{\mu,0}$ is the Kronecker delta, $\alpha$ is the detuning of the pump laser and $D_\mu$ is the integrated dispersion defined by $D_{\mu} = \nu_{\mu} - (\nu_{0}+\textrm{FSR}\,\mu)$, where $\nu_{\mu}$ is the cold cavity resonance frequency of mode $\mu$ and FSR is the free spectral range of the resonator \cite{kippenberg2018dissipative, black2022optical}. Both $\alpha$ and $D_{\mu}$ are normalized by the halfwidth of the resonator $\Delta\nu/2=\frac{\kappa}{4\pi}$, and the time $t$ is normalized by $\frac{2}{\kappa}$, where $\kappa$ is the overall loss rate of the ring. 

The interaction between the CW and CCW pump mode inside the PhCR due to coherent backscattering induced by the nanostructured inner-wall modulation is characterized by $\xi$. For a PhCR without a pump reflector, $\xi$ is approximately equal to the pump mode frequency split normalized by $\Delta\nu/2$ when this split is much larger than $\Delta\nu/2$; see the Supplemental Material. While the PhCR provides universal phase matching, the CE depends greatly on the linear coupling dynamics \cite{black2022optical}. The overall loss rate can be divided into two parts: $\kappa=\kappa_i+\kappa_c$, where $\kappa_i$ is the intrinsic loss rate while $\kappa_c$ is the rate of energy exchange between the bus waveguide and the ring. The ratio between them $K=\kappa_c/\kappa_i$ is called the coupling coefficient. It is essential to CE since when $K$ is high, more light is coupled out with the same energy dissipation in the resonator. Due to this coupling, there is a difference between the fields on the input and output side of the PhCR in the bus waveguide. Moreover, due to the presence of the pump reflector, the input and output to the resonator are counterpropagating in relation to the transmitted and reflected directions. Therefore, we create a notation here to clarify this difference. The fields with superscript i and o in Fig. 1(a) denote input and output fields relative to the PhCR.  
For the transmitted direction, the field input to resonator $E_{t\mu}^i$ and coming out of the resonator $E_{t\mu}^o$ are
\begin{equation}
\begin{split}
E_{t\mu}^i &= \sqrt{\frac{K+1}{2K}}F_t\delta_{\mu,0}\\ E_{t\mu}^o &= E_{t\mu}^i - \sqrt{\frac{2K}{K+1}}E_{t\mu} = \sqrt{\frac{K+1}{2K}}(F_t\delta_{\mu,0}-r_{\textrm{EF}}E_{t\mu})\\r_{\textrm{EF}} &= \frac{2K}{K+1},
\end{split}
\label{coupling theory}
\end{equation}
where $r_{\text{EF}}$ is a conversion coefficient between the field in the bus waveguide and the normalized driving force $F_t$ or $F_r$ inside the resonator, and is of utmost importance to optimize CE. The proof of these formulas can be found in the Supplemental Material.

 Due to the PhC, there is a CCW propagating pump mode inside the ring. With the addition of a pump reflector in the bus waveguide, the pump mode field $E_{t0}^o$ is reflected back, becoming $E_{r0}^i = rE_{t0}^o$, further converted into the CCW driving force $F_r = r(F_t-r_{\textrm{EF}}E_{t0})$, and thus reduces the waste and improves the CE. We assume the pump reflector only reflects the pump frequency. Here $r$ is the reflection coefficient for pump mode of the reflector $r = \sqrt{R}e^{i\Phi}$, where $R$ is the reflectivity and $\Phi$ the reflector phase. The reflected wave inside the bus waveguide can be similarly written as: $E_{r\mu}^i = \sqrt{\frac{K+1}{2K}}F_r\delta_{\mu,0}$ and $E_{r\mu}^o = E_{r\mu}^i - \sqrt{\frac{2K}{K+1}}E_{r\mu}$. Since the input of the whole system is proportional to $|F_t|^2$, we will replace $F_t$ with $F$ and assume it to be a real number for convenience in the text below. Then, the input power $P$ of the system can be written as
\begin{equation}
\begin{split}
P = \eta F^2
\end{split}
\label{eta}
\end{equation}
 where $\eta$ is the conversion coefficient between P and $F^2$, and originates from the normalization of the modified LLE (Eqn (\ref{LLE})). Therefore, $\eta$ depends on the halfwidth of the resonator which is related to $K$, the linear and nonlinear refractive index of the PhCR and the mode volume \cite{PhysRevApplied.14.014006}. 

With the preparation above, we derive the CE formula for a PhCR with pump reflector. The derivation is based on energy conservation, so we study the energy flow of the system. The input field is $E_{t0}^i = \sqrt{\frac{K+1}{2K}}F$. For the output, we investigate the power in the pump mode and other modes separately. Consider first the pump mode. The transmitted wave is $\sqrt{1-|r|^2}E_{t0}^o = \sqrt{1-|r|^2}\sqrt{\frac{K+1}{2K}}(F-r_{\textrm{EF}}E_{t0})$, ignoring the phase, which is not related to the intensity. The reflected wave is $E_{r0}^o = \sqrt{\frac{K+1}{2K}}(r(F-r_{\textrm{EF}}E_{t0})-r_{\textrm{EF}}E_{r0})$. For non-pump modes $\mu \neq 0$, the driving forces and the pump reflector have no effect. Therefore, the transmitted and reflected fields are $E_{t\mu}^o = -\sqrt{\frac{2K}{K+1}}E_{t\mu}$ and $E_{r\mu}^o = -\sqrt{\frac{2K}{K+1}}E_{r\mu}$, and their total power equals $r_{\textrm{EF}}\sum_{\mu\neq0}(|E_{t\mu}|^2+|E_{r\mu}|^2)\frac{\omega_{\mu}}{\omega_{0}}$, where $\omega_{\mu}$ denotes the measured angular frequency of mode $\mu$. Note that for four-wave mixing, the idler (with mode number $\mu_{i}$) and signal (with mode number $\mu_{s}$) are generated in pairs and their angular frequencies satisfy $\omega_{\mu_{i}}+\omega_{\mu_{s}}=2\omega_{0}$. The total power can be further simplified to $r_{\textrm{EF}}I_c$, where $I_c = \sum_{\mu\neq0}(|E_{t\mu}|^2+|E_{r\mu}|^2)$. Due to energy conservation, for the steady state, the input power equals the total power in both directions plus the intrinsic loss, which we write as $\sum_\mu\frac{2}{K+1}(I_{t\mu}+I_{r\mu})$, according to definition of $K$. The factor 2 arises due to the loss normalization to the half linewidth. Then, the energy conservation equation becomes
\begin{equation}
\begin{split}
|E_{t0}^i|^2 =& |\sqrt{1-|r|^2}E_{t0}^o|^2 + |E_{r0}^o|^2 + r_{\textrm{EF}}I_c \\&+ \sum_\mu\frac{2}{K+1}(I_{t\mu}+I_{r\mu}).
\end{split}
\end{equation}
The solution of this equation is 
\begin{equation}
\begin{split}
I_c = F\textrm{Re}[E_{t0}+r^*E_{r0}] - (I_{t0}+I_{r0}+r_{\textrm{EF}}\textrm{Re}[rE_{t0}E_{r0}^*]).
\end{split}
\end{equation}
According to the definition of CE,
\begin{equation}
\begin{split}
\textrm{CE} = \frac{r_{\textrm{EF}}I_c}{|E_{t0}^i|^2} =  {r_{\textrm{EF}}}^2(&\frac{\textrm{Re}[E_{t0}+r^*E_{r0}]}{F}\\-&\frac{I_{t0}+I_{r0}+r_{\textrm{EF}}\textrm{Re}[rE_{t0}E_{r0}^*]}{F^2}).
\end{split}
\label{general CE}
\end{equation}
We further simplify the formula above for some specific cases. For an ordinary resonator without pump reflector ($E_{r0}=0$), and equation (\ref{general CE}) becomes $\textrm{CE} = (\frac{2K}{K+1})^2(\sqrt{\frac{I_{t0}}{F^2}}-\frac{I_{t0}}{F^2})$ \cite{sayson2019octave,stone2022conversion}. However, PhCR are quite different. To achieve wide span OPO with normal GVD, the mode split is much larger than the halfwidth $\xi\gg1$, which we call the large mode split approximation (see Supplemental Material). The strong interaction between the CW and CCW wave of the pump mode establishes coherence between them for the red-shifted resonance mode ($\alpha>0$), hence $E_{t0}\approx-E_{r0}$, which suggests a standing wave inside the resonator; see Supplemental Material. Then, equation (\ref{general CE}) reduces to
\begin{equation}
\begin{split}
\textrm{CE} &= {r_{\textrm{EF}}}^2(\frac{\textrm{Re}[E_{t0}(1-r^*)]}{F}-\frac{I_{t0}(2-r_{\textrm{EF}}\textrm{Re}[r])}{F^2}) \\&\leq {r_{\textrm{EF}}}^2(\sqrt{\frac{I_{t0}}{F^2}}|1-r|-\frac{I_{t0}}{F^2}(2-r_{\textrm{EF}}\textrm{Re}[r])).
\end{split}
\end{equation}
On the other hand, the $I_{t0}$ when OPO exists is actually determined by the phase matching condition \cite{sayson2019octave}. According to CE formula above, a smaller $I_{t0}$ enables the device to achieve the same CE at smaller $F$. The minimum of $I_{t0}$ is 1 and the inequality will become equality when $E_{t0}$ has the same complex angle as $1-r$. This can be satisfied by optimizing $\xi$ and sweeping $\alpha$, which we call optimal phase matching (OPhM); see the Supplemental Material. Then, CE reduces to
\begin{equation}
\begin{split}
\textrm{CE} = {r_{\textrm{EF}}}^2(\frac{1}{F}|1-r|-\frac{1}{F^2}(2-r_{\textrm{EF}}\textrm{Re}[r])).
\end{split}
\label{simplified CE}
\end{equation}
Further, $F$ at threshold is the solution of $\textrm{CE}=0$, which we write as
\begin{equation}
\begin{split}
F_{\textrm{thre}} = \frac{2-r_{\textrm{EF}}\textrm{Re}[r]}{|1-r|}.
\end{split}
\label{Fthre}
\end{equation}
The threshold power in the experiment $P_{\textrm{thre}}$ can be expressed as
\begin{equation}
\begin{split}
P_{\textrm{thre}} = \eta{F_{\textrm{thre}}}^2.
\end{split}
\label{Pthre}
\end{equation}
If we combine equation (\ref{Pthre}) with equation (\ref{eta}), we can cancel $\eta$ and express $F$ as
\begin{equation}
    \begin{split}
    F = F_{\textrm{thre}}\sqrt{\frac{P}{P_{\textrm{thre}}}}
    \end{split}
    \label{Fcontrol}
\end{equation}
by which we can control $F$ in our experiment through the input optical power. Equation (\ref{simplified CE}) depends on 4 variables: $F$, $\Phi$, $R$ and $K$ (included in $r_{\textrm{EF}}$). When we increase input power, CE increases first, and then it drops. The maximum of CE happens at $F=2F_{\textrm{thre}}$ (namely $P=4P_{\textrm{thre}}$), which we call saturation CE and denote by $ \textrm{CE}_{\textrm{sat}}$:
\begin{equation}
\begin{split}
\textrm{CE}_{\textrm{sat}}=\frac{{r_{\textrm{EF}}}^2}{4}\frac{|1-r|^2}{2-r_{\textrm{EF}}\textrm{Re}[r]}.
\end{split}
\label{CEsat}
\end{equation}
For a PhCR without a pump reflector, $F_{\textrm{thre}}=2$ and $ \textrm{CE}_{\textrm{sat}}=\frac{1}{2}(\frac{K}{K+1})^2$. For the fixed reflectivity $R$, $F_{\textrm{thre}}$ can be smaller than 2 at some $\Phi$, and $ \textrm{CE}_{\textrm{sat}}$ is maximized when $\Phi=\pi$. The maximum of $ \textrm{CE}_{\textrm{sat}}$ is $\frac{K^2}{(K+1)(K+1/2)}$ when $r = -1$, which surprisingly is even greater than the CE limit of ordinary resonators.

\paragraph*{Experiment.}
\begin{figure}[t] \centering \includegraphics[width=0.45\textwidth]{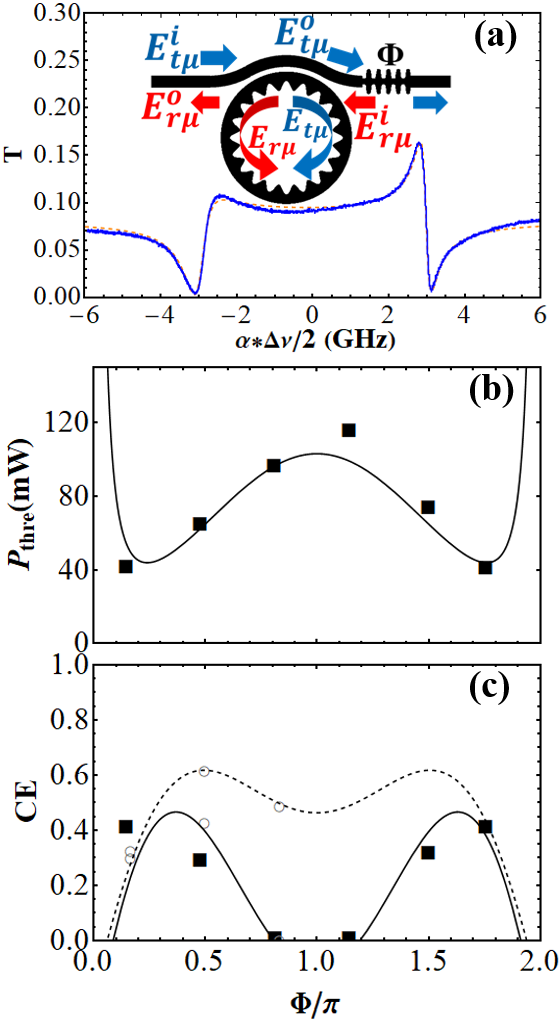}
\caption{\textbf{$\Phi$ dependence} (a) A diagram and a transmission trace of a PhCR with pump reflector. The blue curve is the normalized transmittance from experiment. The orange dashed line plots the fit. (b) $P_{\textrm{thre}}$ vs $\Phi$ with $K=6.0\pm0.6$, $R=0.8\pm0.1$. The solid curve shows analytic $P_{\textrm{thre}}$ with $\eta = 29.7$ mW and the squares are experimental data. (c) CE vs $\Phi$ with $K=6.0\pm0.6$, $R=0.8\pm0.1$. The solid and dashed curve plot the analytic CE at OPhM for $F=$1.8 and 2.3 respectively. Empty circles are simulation results with same conditions as curves. Squares are experimental results with $F=1.80\pm0.05$.}
\label{fig1}
\end{figure}


We explore and measure OPO laser converters with a PhCR and a pump reflector, demonstrating the close connection between lasers we implement with integrated photonics and our predictive model of the system. Hence, our work opens up a robust platform to realize exceptionally designable laser converters with high efficiency and access to broad wavelength bands. Figure 1 presents the foundations of our system, including the $\Phi$ dependence of threshold power and CE. The pump reflector has a structure of a series of teeth with tapering period and amplitude, which controls the width and the location of the reflection band. 
We model the reflector using finite element method software. The PhCR enables deterministic single mode splitting by periodically modulating the resonator inner-wall with a nanostructured-period that determines the split resonant wavevector. The amplitude of the modulation is proportional to 
$\xi$ \cite{black2022optical}. 
We implement the PhCR with pump reflector by use of the tantala integrated photonics platform that we have developed \cite{jung2021tantala, black2021group}. Tantala films of 570 nm thickness are deposited by a commercial vendor, FiveNine Optics, on a thermally oxidized silicon wafer, and we realize our PhCR and pump reflector designs via electron-beam lithography and a fluorine inductively coupled plasma-reaction ion etch. Our 2-day fabrication period yields >40 chips with ~100 resonators per chip and includes an overnight 500$^\circ$C thermal anneal in air to reduce oxygen vacancies in the tantala film. We test our devices by use of a widely tunable external cavity diode laser 1550 nm that is further amplified with an erbium-doped fiber. We convey pump light to the device by lensed optical fibers, which couple light into and out of our chips using an on-chip inverse taper for optimal mode-matching to the fibers.


To test our analytic model, we compare the results with experiment and numerical simulation using equation (\ref{LLE}). By measuring the loss between the lensed fibers and our chips, we infer the on-chip power from the output power measured by an optical spectrum analyzer. Additionally, we use a circulator before the device to collect the reflected light; see Supplemental Material.
We calculate reflectivity from the ratio between the reflected spectrum of devices with pump reflectors and the transmitted spectrum of devices without reflector. We find the measured reflectivity varies from 72\% to 83\% due to the uncertainty of the measurement and the tolerance of fabrication between different chips. By rotating the nanostructured inner-wall resonator modulation, we equivalently change the relative phase $\Phi$ between reflected light in the bus waveguide and the field inside the PhCR. We measure $\Phi$ as well as $K$ and $\xi$ 
by fitting the transmission, $T$; see Fig. 1(a) and the Supplemental Material. 
Figure 1(b) shows the $\Phi$ dependent measurement of $P_{\textrm{thre}}$ (black squares; Fig. 1(b)), which is consistent with equation (\ref{Fthre}) and (\ref{Pthre}) (solid curve; Fig. 1(b)) after fitting the conversion coefficient, $\eta = 30$ mW. 

Figure 1(c) shows the $\Phi$ dependent measurement of CE  (black squares; Fig. 1(c)). We control $F$ to be 1.80 $\pm$ 0.05 by comparing the input power with $P_{\textrm{thre}}$ for each device, according to equation (\ref{Fcontrol}). 
We calculate CE from the inferred on-chip output and input power, with an uncertainty of 0.4 dB. The solid and dashed curves are the analytic CE (equation (\ref{simplified CE})) at OPhM with $F=$ 1.8 and 2.3, respectively. 
Also, we perform a numerical simulation based on equation (\ref{LLE}) to verify our analytic CE formula (grey empty circles in Fig. 1(c)). Both simulation and experiment results match our analytical CE model. Figure 1 demonstrates that the behavior of the PhCR is sensitive to $\Phi$, which is due to the strong coherence between $E_{t0}$ and $E_{r0}$ when $\xi\gg1$ and the phase shift that the pump reflector adds to $F_r$. Although the device with $\Phi=\pi$ reflector has the maximum saturation CE, for small input power, the devices with reflectors of other phases achieved higher CE first due to lower $F_{\textrm{thre}}$. 

\begin{figure}[t] \centering \includegraphics[width=0.45\textwidth]{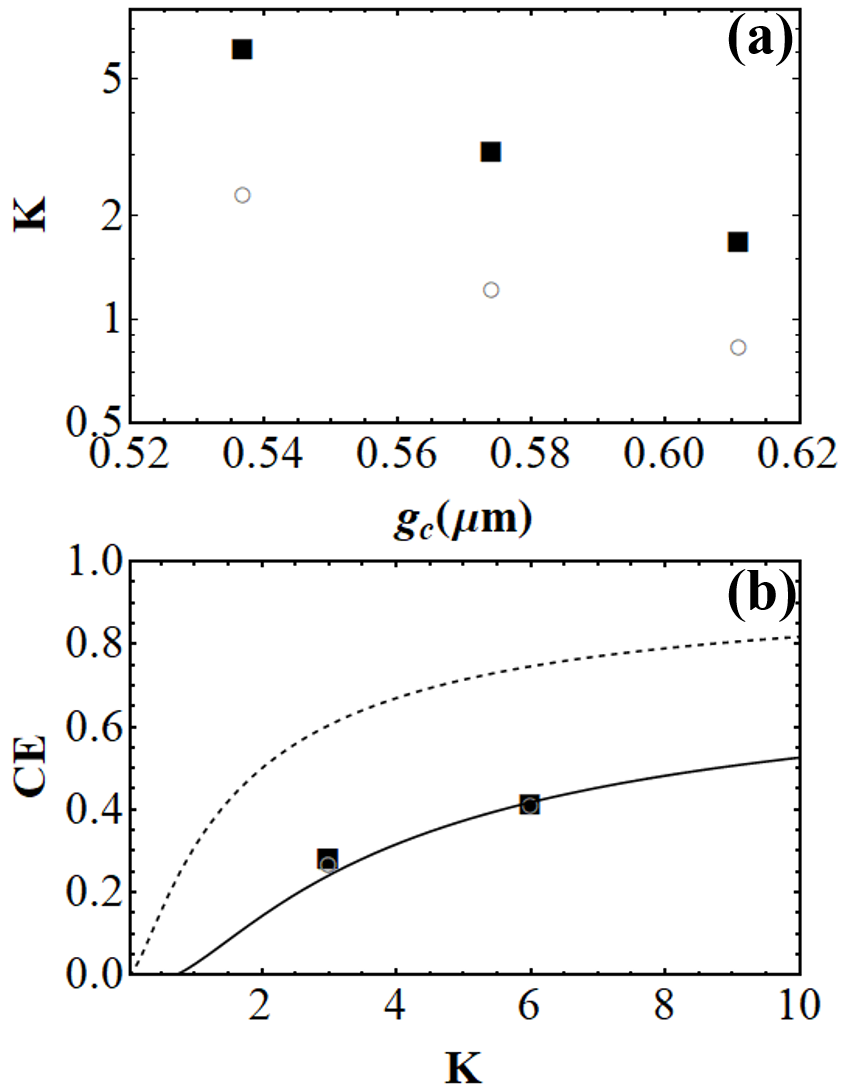}
\caption{\textbf{$K$ dependence} (a) $K$ vs $g_c$. The black squares are the measured $K$ while the empty circles are simulation results using Lumerical FDTD. (b) CE vs $K$. The solid curve is analytic CE at OPhM with $F=1.85$, $\Phi=1.715\pi$, $R=0.8$. The dashed curve is the analytic saturation CE with $\Phi=\pi$, $R=0.8$. The squares are experimental results with approximately same condition. The empty circles are the simulation results with the same parameters.}
\label{fig2}
\end{figure}

The coupling coefficient $K$ is the critical factor to increase CE of OPO extracted from the resonator. We experimentally vary this parameter by adjusting the gap, $g_c$, between the bus waveguide and the PhCR. Figure 2(a) shows the dependence of $K$ on $g_c$. The black squares are experimental results and the empty circles are simulation results, using a finite element method software. They show exponential dependence of $K$ on $g_c$ with a shift in prefactor, which is likely due to the tolerance or incomplete etch of gaps in fabrication. Figure 2(b) shows the dependence of CE on $K$. The solid curve plots the analytic CE at OPhM with $F=1.85$, $\Phi=1.715\pi$ and $R = 0.8$. The black squares and the empty circles are the corresponding measured results and the simulation results with the same parameters obtained from the modified LLE equation (\ref{LLE}). The results are nearly overlapping, demonstrating the validity of the derived analytic expressions. The dashed curve is the analytic $ \textrm{CE}_{\textrm{sat}}$ with the same reflectivity but optimal phase $\Phi=\pi$. It shows that even without modification on the reflector structure, CE can reach nearly 80\% with $K=6$ when $\Phi$ is optimized and the input power is increased to saturation, $P=4P_{\textrm{thre}}$. In addition, according to equation (\ref{CEsat}), with an ideal reflectivity $R=100\%$, CE can reach the theoretical upper boundary $\frac{K^2}{(K+1)(K+1/2)}$, which asymptotically approaches unity when $K\rightarrow+\infty$. 

\begin{figure}[t] \centering \includegraphics[width=0.45\textwidth]{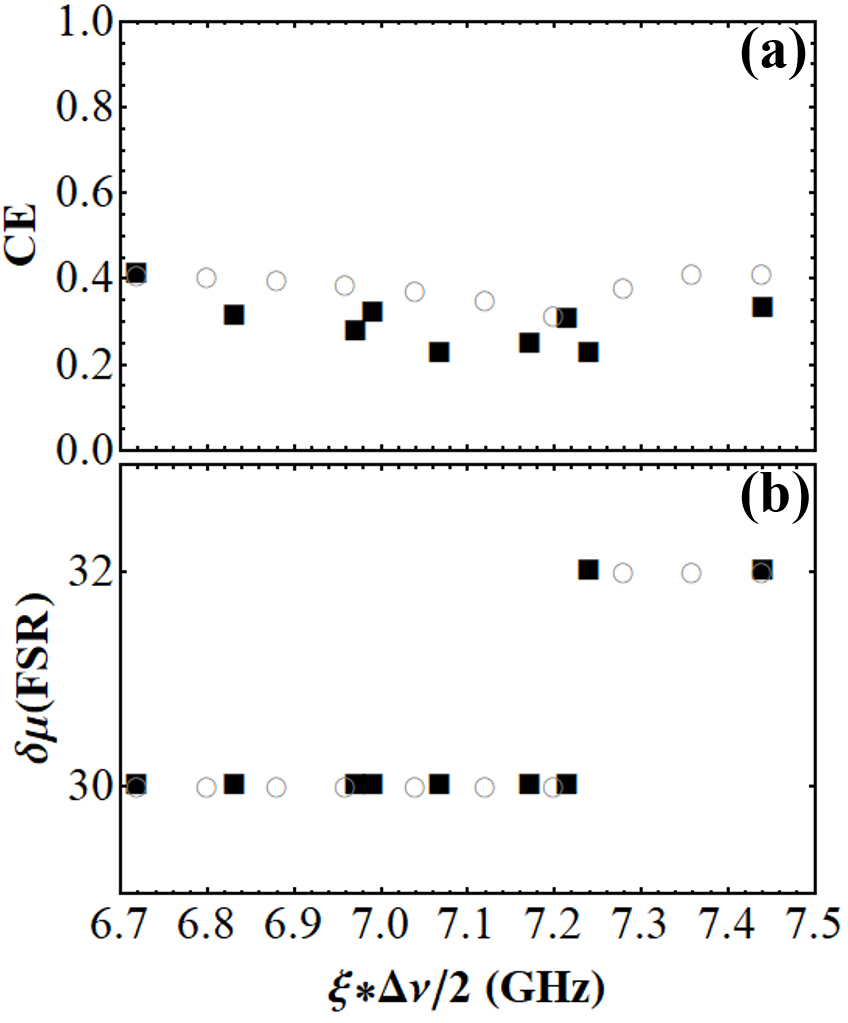}
\caption{\textbf{$\xi$ dependence} (a), (b) CE vs $\xi$ and $\delta\mu$ vs $\xi$. The black squares are the experiment results and the empty circles are the simulation results.}
\label{fig3}
\end{figure}

We explore the CE dependence on OPhM and the corresponding output signal and idler frequencies that are generated. As stated previously, 
our simplified CE formula (equation (\ref{simplified CE})) is valid only when OPhM is satisfied. In order to achieve OPhM, we sweep $\xi$ by adjusting the modulation amplitude of the PhCR's nanostructured inner-wall. This leaves the broader dispersion unchanged, while enabling different phase-matching conditions for the pump mode split by $\xi$. Figure 3(a) shows the CE dependence of $\xi$ with all other parameters fixed. We find CE spans $\approx$20 - 40 \%, where the experimental results (black squares) are in good agreement with our numerical simulation based on equation (\ref{LLE}) (grey circles). The small decrease in CE corresponds to a change in the generated signal and idler frequencies, whose span in FSR is $\delta\mu$, as seen in Fig. 3(b). The jump in $\delta\mu$ denotes that OPhM is satisfied with a new pair of PhCR modes, thereby enabling tunable output frequencies through OPO laser conversion while maintaining high CE. 


\begin{figure}[t] \centering \includegraphics[width=0.43\textwidth]{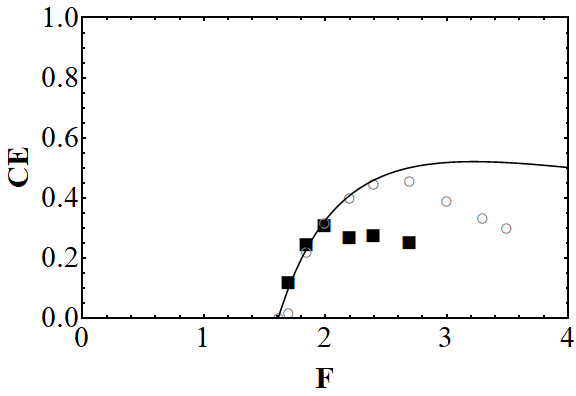}
\caption{\textbf{$F$ dependence} The solid curve and empty circles shows the analytic CE and simulation results with $R=0.72$, $\Phi=1.372\pi$, $K=2.8$, respectively. The black squares are experiment results with $R=0.7\pm0.1$, $\Phi=(1.36\pm0.05)\pi$, $K=2.8\pm0.3$.}
\label{fig4}
\end{figure}

We also investigate the dependence of CE on F, the pump driving field which is controlled by the input pump optical power. 
Figure 4 summarizes our results. The solid curve and empty circles are the analytic CE (equation (\ref{simplified CE})) and simulation results at OPhM with $R=0.72$, $\Phi=1.372\pi$, $K=2.8$. 
The black squares are the experimental results with near parameters. The results are consistent at small $F$, but diverge at larger $F$. We attribute this divergence to two reasons. First, when $F$ is large, more than one pair of modes can achieve phase matching, which we call cascade, and decentralize the pump mode power. Second, the imperfect environment in the experiment is not fully described by either our analytical model or numerical simulation. The reflectors in our devices actually have some minor effect on the signal and idler modes and 
$K$ at the idler mode is larger than 
$K$ at the pump and signal modes. When CE is high, the reflected idler light by the reflector might excite other four wave mixing processes, further decentralizing the pump mode power. These imperfections might be further improved by incorporating higher-order dispersion terms to the simulation and by mitigating higher-order mode crossings, which disrupt the ideal dispersion. The reflectivity on the signal and idler can be reduced with a better reflector. The wavelength dependence of K can be suppressed by redesigning the width of bus waveguide and the length of its 
coupling region (see the diagram in Fig. 1(a)). Moreover, cascading can be avoided 
by utilizing other dispersion regimes and values of $\xi$, or operating with a larger FSR. However, as a trade-off, a finer sweep of $\xi$ will be required in order to reach OPhM at the targeted modes.

\section*{Discussion and conclusion}
In this article, we show that a pump reflector can reduce OPO threshold power and increase CE by controlling the coupling of counterpropagating pump lasers. We develop a theory to explain this mechanism and derive analytic formulas of CE and threshold power for 
nanostructured-resonator OPO. Our analytical model has been systematically verified in our experiments and we have obtained $(41\pm4)$\% CE and $(40\pm4)$ mW on-chip power for OPO laser-conversion. Higher CE 
is limited by the non-ideal reflector spectrum and cascade in our experiment
Through nanophotonic design, we highlight a regime of OPO with high CE and universal phase matching and realize a robust laser platform. 

\begin{acknowledgments}
We thank Yan Jin and Charles McLemore for carefully reviewing the manuscript. This research has been funded by NIST, the DARPA LUMOS programs as well as AFOSR FA9550-20-1-0004 Project Number 19RT1019 and NSF Quantum Leap Challenge Institute Award OMA - 2016244.  This work is a contribution of the US Government and is not subject to copyright. Mention of specific companies or trade names is for scientific communication only and does not constitute an endorsement by NIST. 
\end{acknowledgments}

\section*{Supplemental material}
\maketitle

\subsection{Coupling theory}
Here we derive equation (2) in the main text. Since the coupler is not related to 
the PhCR 
or the pump reflector, we can limit the derivation in some simplified cases. We omit subscript t and r below since the theory is the same for both directions. Notice that $\frac{\partial I_\mu}{\partial t} = 2\textrm{Re}[E_\mu^*\frac{\partial E_\mu}{\partial t}]$. 
For each mode and direction, according to equation (1) in the main text, $\frac{\partial I_\mu}{\partial t} = -2I_\mu$. Due to the definition of $K$, the photon flux that goes into the waveguide is $\frac{2K}{K+1}I_\mu$, so $E_\mu^o=e^{i\gamma}\sqrt{\frac{2K}{K+1}}E_\mu$, where $\gamma$ denotes the phase difference. 
Adding the driving force $F\delta_{\mu,0}$, $E_\mu^i$ can be denoted by $xF\delta_{\mu,0}$ where $x$ is an unknown coefficient and $E_\mu^o=e^{i\gamma}\sqrt{\frac{2K}{K+1}}E_\mu + e^{i\epsilon}xF\delta_{\mu,0}$ due to superposition, where $\epsilon$ denotes another unknown phase. The power that goes from the bus waveguide to the resonator is $|E_\mu^i|^2-|E_\mu^o|^2 = -2\textrm{Re}[e^{i(\epsilon-\gamma)}\sqrt{\frac{2K}{K+1}}E_\mu^*xF\delta_{\mu,0}]-\frac{2K}{K+1}I_\mu$. On the other hand, from equation (1) in the main text, we have  $\frac{\partial I_\mu}{\partial t}=-(\frac{2}{K+1}+\frac{2K}{K+1})I_\mu + 2\textrm{Re}[F\delta_{\mu,0}E_\mu^*]$. Due to energy conservation,  $\frac{\partial I_\mu}{\partial t} = |E_\mu^i|^2 - |E_\mu^o|^2 - \frac{2}{K+1}I_\mu$, where the last term is due to intrinsic loss. This should 
hold for arbitrary $E_\mu$, so $x=-e^{i(-\epsilon+\gamma)}\sqrt{\frac{K+1}{2K}}$. Since only magnitude of $E_\mu^i$ and $E_\mu^o$ can be measured, we can add phases $-e^{i(\epsilon-\gamma)}$ and $-e^{-i\gamma}$ and 
rewrite them as $E_{\mu}^i = \sqrt{\frac{K+1}{2K}}F\delta_{\mu,0}$, $E_{\mu}^o = \sqrt{\frac{K+1}{2K}}(F\delta_{\mu,0}-r_{\textrm{EF}}E_{\mu})$,$r_{\textrm{EF}} = \frac{2K}{K+1}$.

\subsection{Large mode split approximation}
Here we prove $E_{t0} \approx -E_{r0}$ with large mode split approximation for the red resonance mode ($\alpha > 0$). For steady state, the equation (1) in the main text of pump mode in both directions can be written as.
\begin{equation}
\begin{split}
0 = -&(1+i\alpha)E_{t0}-i\frac{\xi}{2}E_{r0}\\+&  i(\sum_{\mu_1,\mu_2}E_{t\mu_1}E_{t\mu_2}E_{t(\mu_1+\mu_2)}^*+2E_{t0}\sum_{\mu_3}I_{r\mu_3}) + F_t\\
0 = -&(1+i\alpha)E_{r0}-i\frac{\xi}{2}E_{t0}\\+&  i(\sum_{\mu_1,\mu_2}E_{r\mu_1}E_{r\mu_2}E_{r(\mu_1+\mu_2)}^*+2E_{r0}\sum_{\mu_3}I_{t\mu_3}) + F_r.
\end{split}
\label{steady pump}
\end{equation}
Summing 
these two equalities, we have
\begin{equation}
\begin{split}
0 = -&(1+i(\alpha+\frac{\xi}{2}))(E_{t0}+E_{r0}) \\+&  i(\sum_{\mu_1,\mu_2}(E_{t\mu_1}E_{t\mu_2}E_{t(\mu_1+\mu_2)}^*+E_{r\mu_1}E_{r\mu_2}E_{r(\mu_1+\mu_2)}^*) \\+& 2E_{t0}\sum_{\mu_3}I_{r\mu_3}+2E_{r0}\sum_{\mu_3}I_{t\mu_3}) + F_t + F_r.
\end{split}
\end{equation}
Notice that $F_t$,$F_r$,$E_{t0}$,$E_{r0}\sim1$ and the magnitudes for the other modes are even smaller. So, $E_{t0} = -E_{r0} + O(1/\xi)$ when $\xi\gg1$. The $O()$ here is used to describe its asymptotic behavior. 
With this conclusion and equation (1) in the main text for the transmitted pump mode, we further obtain:
\begin{equation}
\begin{split}
0 = -&(1+i(\alpha-\frac{\xi}{2}))E_{t0}+O(1)\\+&  i(\sum_{\mu_1,\mu_2}E_{t\mu_1}E_{t\mu_2}E_{t(\mu_1+\mu_2)}^* + 2E_{t0}\sum_{\mu_3}I_{r\mu_3}) + F_t.
\end{split}
\label{forward}
\end{equation}
Therefore, $E_{t0}$ is small unless $\alpha = \xi/2(1+O(1/\xi))$, which suggests $\alpha\gg1$ as well. Besides, it can be easily obtained from symmetry that the blue resonance mode ($\alpha < 0$) happens at $\alpha = -\xi/2(1+O(1/\xi))$ and the field satisfies $E_{t0} = E_{r0} + O(1/\xi)$. Thus, $\xi$ is approximately equal to the normalized mode split in this case.

\subsection{Optimal phase matching (OPhM)}
Here, we give an argument that OPhM can be achieved by sweeping $\alpha$ and $\xi$. We first prove the minimum of $I_{t0}$ when OPO exists is 1. Since the higher order OPO usually has a much smaller intensity that the first order OPO, we can assume there are only three modes: idler, pump and signal, with mode number $-\mu$, 0, and $\mu$ respectively. Then, equation (1) in the main text for the CW idler and signal modes can be simplified to:
\begin{equation}
\begin{split}
\frac{\partial E_{t\mu}}{\partial t} = &-(1+i(\alpha+D_\mu+I_{t\mu}-2I))E_{t\mu} \\&+ i{E_{t0}}^2E_{t(-\mu)}^*\\
\frac{\partial E_{t(-\mu)}^*}{\partial t} = &-(1-i(\alpha+D_{-\mu}+I_{t(-\mu)}-2I))E_{t(-\mu)}^* \\&- i{E_{t0}^*}^2E_{t\mu} 
\end{split}
\label{parametric oscillation}
\end{equation}
where $I=\sum_{\mu_3}(I_{t\mu_3}+I_{r\mu_3})$ is the total intensity in the ring. The eigenvalue with a larger real part of this differential equation is
\begin{equation}
\begin{split}
\lambda = &-1\\&+\sqrt{{I_{t0}}^2-(\alpha+\frac{D_\mu+D_{-\mu}}{2}+\frac{I_{t\mu}+I_{t(-\mu)}}{2}-2I)^2}\\
&-i\frac{D_\mu-D_{-\mu}+I_{t\mu}-I_{t(-\mu)}}{2}.
\end{split}
\end{equation}
For steady state of OPO, the real part of $\lambda$ is zero, which suggests
\begin{equation}
I_{t0} = \sqrt{1+(\alpha+\frac{D_\mu+D_{-\mu}}{2}+\frac{I_{t\mu}+I_{t(-\mu)}}{2}-2I)^2}.
\label{forward pump intensity}
\end{equation}
Its minimum occurs when $\alpha$ equals
\begin{equation}
\alpha_t^{\textrm{opt}} = -\frac{D_\mu+D_{-\mu}}{2}-\frac{I_{t\mu}+I_{t(-\mu)}}{2}+2I .
\end{equation}
The eigensolution of equation (\ref{parametric oscillation}) satisfies:
\begin{equation}
\begin{split}
&I_{t\mu}(-1 + i(\alpha - \alpha_t^{\textrm{opt}})) = i{E_{t0}^*}^2E_{t\mu}E_{t(-\mu)}\\
&I_{t\mu}=I_{t(-\mu)}.
\end{split}
\label{eigensolution}
\end{equation}
The conclusions above can also be applied to the reflected direction:
\begin{equation}
\begin{split}
&I_{r0} = \sqrt{1+(\alpha+\frac{D_\mu+D_{-\mu}}{2}+\frac{I_{r\mu}+I_{r(-\mu)}}{2}-2I)^2}\\
&\alpha_r^{\textrm{opt}} = -\frac{D_\mu+D_{-\mu}}{2}-\frac{I_{r\mu}+I_{r(-\mu)}}{2}+2I\\
&I_{r\mu}(-1 + i(\alpha - \alpha_r^{\textrm{opt}})) = i{E_{r0}^*}^2E_{r\mu}E_{r(-\mu)}\\
&I_{r\mu}=I_{r(-\mu)}.
\label{backward pump intensity}
\end{split}
\end{equation}
On the other hand, when $\xi\gg1$, the large mode split approximation yields $I_{t0} \approx I_{r0}$. Since the only difference between equation (\ref{forward pump intensity}) and the first equality in equation (\ref{backward pump intensity}) is the dependence on $I_{t\mu}$ and $I_{r\mu}$, it suggests that OPO can be generated either in only one direction or in both directions with $I_{t\mu} \approx I_{r\mu}$. In our experiment, we 
find that the ratio between the converted power in 
either direction varies 
substantially while the total converted power is relative stable during the detuning sweep. We attribute differences between the analytic model predictions and experiment to chip-facet reflections and the assumption that only pump light is reflected. 

Next, 
we prove that by sweeping $\xi$, we can align the complex angle of $E_{t0}$ with $1 - r$. Ignoring higher order OPO, equation (\ref{steady pump}) can be simplified to:
\begin{equation}
\begin{split}
0 = &-(1+i(\alpha+I_{t0}-2I))E_{t0}-i\frac{\xi}{2}E_{r0} \\&+ 2iE_{t\mu}E_{t(-\mu)}E_{t0}^* + F\\
0 = &-(1+i(\alpha+I_{r0}-2I))E_{r0}-i\frac{\xi}{2}E_{t0} \\&+ 2iE_{r\mu}E_{r(-\mu)}E_{r0}^* + r(F-r_{\textrm{EF}}E_{t0}).
\end{split}
\end{equation}
Combine the above equation with equation (\ref{eigensolution}) and (\ref{backward pump intensity}):
\begin{equation}
\begin{split}
\frac{F}{E_{r0}} = &i\frac{\xi}{2}+\frac{E_{t0}}{E_{r0}}(1+\frac{2I_{t\mu}}{I_{t0}}\\+&i(\alpha+I_{t0}-2I-\frac{2I_{t\mu}}{I_{t0}}(\alpha-\alpha_t^{\textrm{opt}})))\\
\frac{rF}{E_{t0}} = &rr_{\textrm{EF}}+i\frac{\xi}{2}+\frac{E_{r0}}{E_{t0}}(1+\frac{2I_{r\mu}}{I_{r0}}\\+&i(\alpha+I_{r0}-2I-\frac{2I_{r\mu}}{I_{r0}}(\alpha-\alpha_r^{\textrm{opt}}))).
\end{split}
\label{three modes}
\end{equation}
With large mode split approximation, $E_{t0} = -E_{r0}+O({1/\xi})$ and 
\begin{equation}
\begin{split}
&\frac{I_{r0}}{I_{t0}}=(1+O(1/\xi))\\
&\frac{E_{t0}}{E_{r0}}+\frac{E_{r0}}{E_{t0}}=-2+O(1/\xi^{2}).
\end{split}
\end{equation}
Sum 
the two equalities in equation (\ref{three modes}):
\begin{equation}
\begin{split}
\frac{F(1-r)}{E_{t0}} = &O(1/\xi)+2-rr_{\textrm{EF}}+\frac{I_c}{I_{t0}}\\+&2i(\alpha-\frac{\xi}{2}+I_{t0}-2I\\&-\frac{I_{r\mu}(\alpha-\alpha_r^{\textrm{opt}})+I_{t\mu}(\alpha-\alpha_t^{\textrm{opt}})}{I_{t0}}).
\end{split}
\end{equation}
By adjusting $\xi$, we can eliminate the imaginary part in the bracket on the right hand side of equation above, and thus align $E_{t0}$ with $1 - r$.

\subsection{Transmission trace fitting}
As discussed in the main text, the pump reflector phase is measured by fitting the transmission trace. In experiment, the signal $V$ on the oscilloscope over the unnormalized detuning $\Gamma$ can be measured directly as $V(\Gamma)$. $V$ is proportional to $T$ and can be written as $V = CT + B$ where $B$ is the background noise. On the other hand, for low input power, the nonlinearity terms in modified LLE can be ignored, and the steady state equation for pump mode can be written as:
\begin{equation}
\begin{split}
0 = &-(1+i\alpha)E_{t0} + (F-i\frac{\xi}{2}E_{r0})\\
0 = &-(1+i\alpha)E_{r0} + (r(F-r_{\textrm{EF}}E_{t0})-i\frac{\xi}{2}E_{t0}).
\end{split}
\label{cold cavity pump}
\end{equation}
By solving the equation above, $T$ can be expressed as:
\begin{equation}
\begin{split}
T = &(1-R)|\frac{F-r_{\textrm{EF}}E_{t0}}{F}|^2\\
= &(1-R)|\frac{(1+i\alpha)^2+(\xi/2)^2-r_{\textrm{EF}}(1+i\alpha)}{(1+i\alpha)^2+(\xi/2)^2-ir_{\textrm{EF}}r\xi/2}|^2.
\end{split}
\label{analytic T}
\end{equation}
We assume that the devices on the same chip have approximately the same $R$ and $\kappa_i$, and they can be measured in experiment. Then, the unknown valuables in the equation above can be expressed as:
\begin{equation}
\begin{split}
r_{\textrm{EF}} &= \frac{2K}{K+1} = 2(1-\frac{\kappa_i}{2\pi\Delta\nu})\\
r &= Re^{i\Phi}\\
\alpha &= 2\Gamma/\Delta\nu.
\end{split}
\end{equation}
 Then, $V$ can be expressed in another way as $V = CT(\Gamma,\xi,\Delta\nu,\Phi) + B$. By nonlinear fitting the measured $V(\Gamma)$, we obtain the known parameters: $\xi$, $\Delta\nu$, and $\Phi$.

\subsection{Experimental setup and output measurement}
\begin{figure}[t] \centering \includegraphics[width=0.43\textwidth]{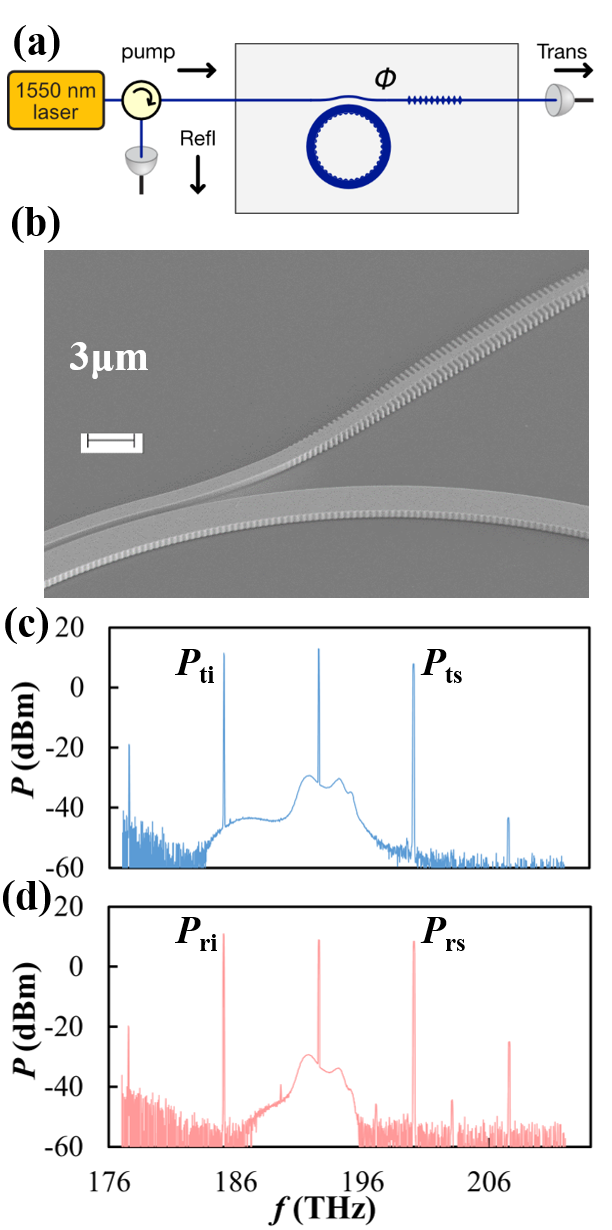}
\caption{(a) A diagram of our experimental setup. (b) A SEM image of a PhCR with pump reflector device. 
(c),(d) The output spectra in transmitted (blue) and reflected (red) directions with $\textrm{CE}=(41\pm4)\%$. The power of idler and signal is labeled by subscripts i and s, and subscripts t and r denote the transmitted and reflected directions, respectively.}
\label{figSM}
\end{figure}
In our experiment, we use a circulator before the device to collect the reflected light, and the output spectra of both directions are measured by the same optical spectrum analyzer (OSA) by using an optical switch. Figure 5 (a) shows a diagram of our setup. Figure 5 (b) is a scanning electron microscope (SEM) image of our PhCR with bus reflector device. The teeth on the ring are the nanostructured side wall modulations, and the teeth on the bus waveguide are the pump reflector.

Figure 5 (c) and (d) show the measured spectra in transmitted (blue), and reflected (red) directions with the highest CE we achieve in experiment ($(41\pm4)\%$). Here, the power of the idler and signal in each direction are labeled by subscript i and s, and the transmitted and reflected directions are labeled by subscript t and r, respectively. The power in these spectra is the on-chip power which is calculated from the off-chip spectra measured by the OSA. 
We calculate CE by
\begin{equation}
    \textrm{CE}=(P_{\textrm{ti}}+P_{\textrm{ts}}+P_{\textrm{ri}}+P_{\textrm{rs}})/P_p,
\end{equation}
where $P_p$ is the on-chip input pump power, which is calculated from the off-chip power measured by a power meter before the circulator.


\clearpage
\bibliography{Haixin} 

\end{document}